\documentclass{ws-ijmpcs}
\begin{document}

\markboth{G. Malfatti {\it et al.}}
{3nPNJL model applied to Proto-neutron Stars Cores}

%
\catchline{}{}{}{}{}
%

\title{Quark-hadron Phase Transition in Proto-Neutron Stars Cores
based on a Non-local NJL Model}

\author{German Malfatti$^{a,b,\dag}$,
G.A. Contrera$^{a,b,c,\ddag}$,
M. Orsaria$^{a,b,\S}$ and
F. Weber$^{d,e,\P}$
\vspace{6px}}

\address{
$^{a}$ CONICET, Godoy Cruz 2290, 1425 Buenos Aires, Argentina.\\
$^{b}$ Grupo de Gravitaci\'on, Astrof\'isica y Cosmolog\'ia,\\
  Facultad de Ciencias Astron{\'o}micas y Geof{\'i}sicas, Universidad Nacional de La Plata,\\
  Paseo del Bosque S/N (1900), La Plata, Argentina.\\
$^{c}$ IFLP, UNLP, CONICET, Facultad de Ciencias Exactas, calle 49 y 115, La Plata, Argentina.\\
$^{d}$ Department of Physics, San Diego State University, 5500 Campanile Drive,\\
  San Diego, CA 92182, USA.\\
$^{e}$ Center for Astrophysics and Space Sciences, University of California,\\
  San Diego, La Jolla, CA 92093, USA.\\
$^{\dag}$germanmalfatti@gmail.com, $^{\ddag}$contrera@fisica.unlp.edu.ar,\\
$^{\S}$morsaria@fcaglp.unlp.edu.ar, $^{\P}$fweber@mail.sdsu.edu
}

\maketitle

\begin{history}
\received{Day Month Year}
\revised{Day Month Year}
\published{Day Month Year}
\end{history}

\begin{abstract}
We study the QCD phase diagram using a non-local SU(3) NJL model with
vector interactions among quarks. We analyze several thermodynamic
quantities such as entropy and specific heat, and study the influence
of vector interactions on the thermodynamic properties of quark
matter. Upon imposing electric charge neutrality and baryon number
conservation on the field equations, we compute models for the
equation of state of the inner cores of proto-neutron stars and providing a non-local treatment of quark matter for
astrophysics.  \keywords{Phase Diagram; Equation of State; Neutron
  Stars.}
\end{abstract}

\ccode{PACS numbers:}

\section{Introduction}	
It is known that quantum chromodynamics (QCD), has two very important
properties, namely asymptotic freedom and confinement. The former
implies that at high momentum transfers, the quarks behave as quasi
free particles, i.e., the interaction between two quarks due to gluon
interchange can be treated using perturbation theory. For this
momentum range, the dispersion processes can be then be calculated
with a very good precision. By contrast, at low momentum exchange
among the quarks ($\lesssim 1$~GeV) QCD is highly nonlinear and leads
to quark confinement. Several approximate methods have been developed
to study the physical processes among quarks in the low momentum range
of QCD. Among them is lattice QCD, which tries to solve the QCD
equations of motion numerically on a discretized space-time
grid\cite{Yndur}. This method, however, presents problems if extended
to finite chemical potentials\cite{Karsch}. Another option is to use
effective QCD models, like the Nambu--Jona-Lasinio model (NJL), which
is described by a Lagrangian which accounts for the main features of
QCD at low energies. The advantage of NJL models is that they can be
extended to finite chemical potentials easily. Local as well as
non-local extensions of the model have been studied in the literature
(see, for instance, Refs. \refcite{{Ranea}}, and references therein).

Non-local extensions of the NJL model at zero temperature have been
used to study hybrid stars with and/or without a quark-hadron mixed
phases in their inner cores (see
Ref.\ \refcite{Spinella,Carvalho,Ranea}, and references
therein). Here, we present an extension of such studies to finite
temperatures in order to explore the role of a quark-hadron phase
transition for proto-neutron stars or core collapse supernovae.

\section{n3PNJL model and phase diagram}

\begin{figure}[pb]
\centerline{\includegraphics[angle=0,width=12cm]{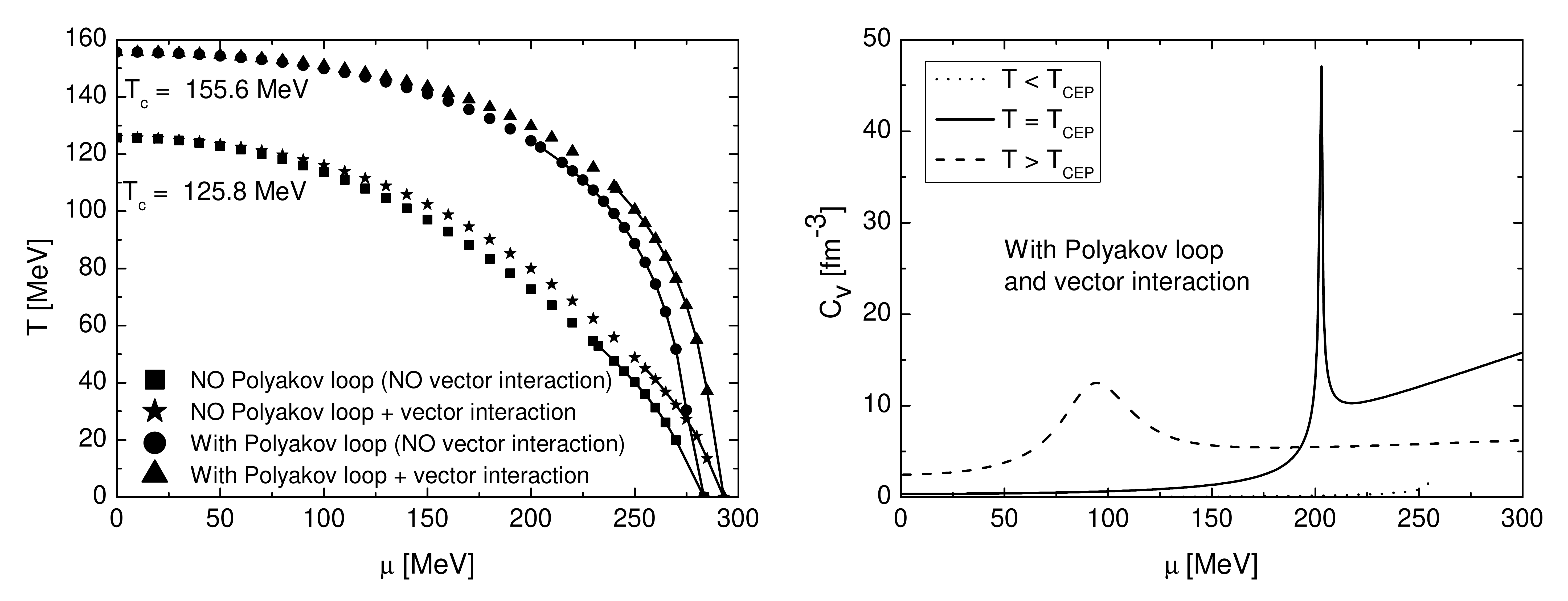}}
\vspace*{8pt}
\caption{Phase diagram of QCD matter. Left panel: temperature, $T$, as
  a function of chemical potential, $\mu$. Right panel: specific heat,
  $C_v$, as a function of $\mu$. (CEP stand for critical end point.)
  The impact of the Polyakov loop and vector interactions on $C_v
  (\mu)$ is very small and therefore not shown separately. The
  locations of irst-order phase transitions are shown by solid
  lines.  \label{f1}}
\end{figure}

\begin{figure}[pb]
\centerline{\includegraphics[angle=0,width=12cm]{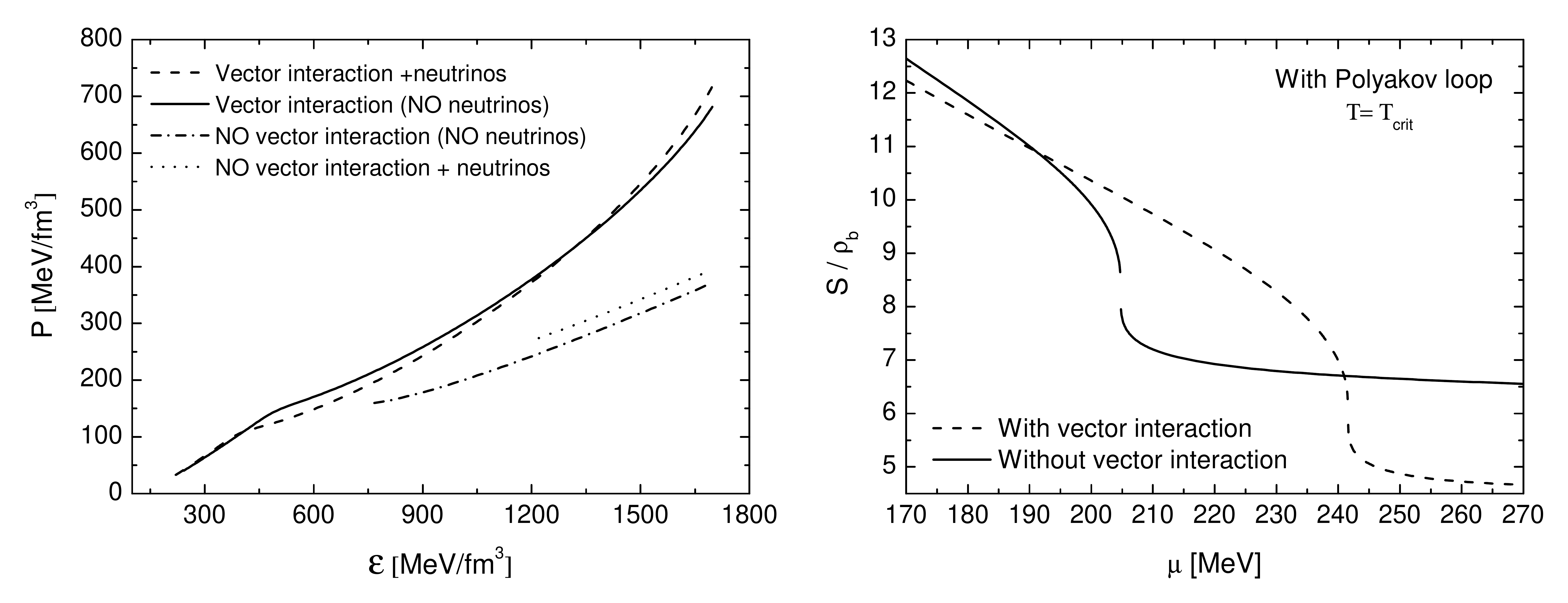}}
\vspace*{8pt}
\caption{Models for the equation of state (pressure versus energy
  density) of proto-neutron star matter (left panel), and entropy per
  baryon (in units of the Boltzmann constant) versus chemical potential
  (right panel). \label{f2}}
\end{figure}

The Euclidean effective action of the model presented in this paper is
for the non-local 3-flavor Polyakov NJL model, including the vector
interactions among quarks. This action is given by
\begin{eqnarray}
S_E &=& \int d^4x \left\{ \bar \psi (x) \left[ -i\gamma_{\mu}D_{\mu} +
  \hat m \right] \psi(x) - \frac{G_s}{2} \left[ j_a^S(x) \ j_a^S(x) +
  j_a^P(x) \ j_a^P(x) \right] \right.  \nonumber \\ & & \qquad \qquad
\left. - \frac{H}{4} \ T_{abc} \left[ j_a^S(x) j_b^S(x) j_c^S(x) -
  3\ j_a^S(x) j_b^P(x) j_c^P(x) \right]\right.  \nonumber \\ & &
\qquad \qquad \left.- \frac{G_{v}}{2} \left[j_{v}^\mu(x)
  j_{v}^\mu(x)\right] + \ {\cal U} \,[{\mathcal A}(x)] \right\}\;,
\label{eq:S_E}
\end{eqnarray}
where $j_a^{S}(x)$, $j_a^{P}(x)$ and $j_{v}^{\mu}(x)$ are the scalar,
pseudoscalar and vector currents, respectively, and
$\mathcal{U}[{\mathcal A}(x)]$ is the effective potential related with
the Polyakov loop. To extend the model to finite temperatures we use
the Matsubara imaginary time formalism\cite{Kapusta}. After the
bosonization of Eq.\ (\ref{eq:S_E}) we obtain the grand canonical
potential for the mean field approximation, from which the phase
diagram (Fig.\ (\ref{f1})) and thermodynamic quantities such as the
specific heat, $C_v$, and specific entropy, $s$, (Fig.\ (\ref{f2}))
can be calculated.

The quark current masses and coupling constants in Eq.\ (\ref{eq:S_E})
can be chosen so as to reproduce the phenomenological values of the
pion decay constant, $f_\pi$, and the meson masses $m_{\pi}$,
$m_\eta$, and $m_{\eta'}$, as described in
Ref.\ \refcite{Contrera:2009hk}. In this work we choose the strange
quark mass to the updated value of $m_s=95$ MeV, with the values of
the other parameters given by $\Lambda = 1071.38$ MeV, $G_s \Lambda^2
= 10.78$, $H \Lambda^5 = -353.29$ and $m_u=m_d=3.63$ MeV. The vector
interaction coupling constant, $G_v$, has been assigned a canonical
value of $G_v = 0.5 G_s$. Finally, we mention that the parameter
$\Lambda$ determines the range of the non-locality in momentum space.

Because of their short mean-free paths, the neutrinos produced in the
core of a proto-neutron star are prevented from leaving the star on a
dynamical time scale.  The number of lepton-to-baryon ratio of such
matter is around $Y_{L_e}\simeq 0.4$, but the exact value depends on
the efficiency of electron capture reactions during the gravitational
collapse of the supernova. The number of muons per baryon is
$Y_{L_{\mu}}=0$, because no muons are present in the stellar matter
prior to the trapping.  On a timescale of 10 to 20 seconds, the
neutrinos diffuse from the star, but leave behind much of their energy
which causes significant heating of the ambient matter\cite{BML}.  The
temperatures generally achieved in the inner 50\% of the stellar core
at the peak of the heating are in the range of 30 to 50 MeV. Following
the heating, the star cools by radiating neutrino--anti-neutrino
pairs, and consequently the temperature drops off to $\sim 1$ MeV
within minutes. A model for the equation of state of such matter,
which describes the inner cores of proto-neutron stars, is shown in
Fig.(\ref{f2}).

\section{Conclusions}

We have constructed models for the equation of state of hot quark
matter in the framework of the n3PNJL model. Knowledge of the equation
of state, a possible critical end point (CEP), and the nature of the
phase transition in the QCD phase diagram itself are critical for a
better understanding of the strongly interacting matter at high
densities and/or temperatures. We have found that quark confinement as
described by the Polyakov loop modifies the phase diagram by
increasing the transitions temperatures. The inclusion of vector
interactions among quarks, on other hand, changes the location of the
CEP. Also, vector interactions stiffen the equation of state, while
the inclusion of neutrinos has the opposite effect (softens the
equation of state).

One of the main motivations of this study concerns the possible
occurrence of quark matter inside of proto-neutron stars. One of the
mechanism for the formation of quark matter in the inner cores of such
stars could be the nucleation of quark matter drops due to the high
pressure environment (high compression) that exists in the centers of
proto-neutron star\cite{Lugones:2016}. Such an investigation, as well
as the effect of neutrino trapping in hot beta-stable quark-hybrid
matter is currently begin carried out.

\section*{Acknowledgments}

G.M., G.A.C.\ and M.O.\ thank CONICET and UNLP for financial
support. M.O.\ acknowledges financial support from American Physical
Society's International Research Travel Award Program. F.W.\ is
supported by the National Science Foundation (USA) under Grant
PHY-1411708.



\begin{thebibliography}{0}    

\bibitem{Yndur} F. J. Yndur{\'a}in, The theory of quarks and gluon
  interactions (Springer-Verlag, Heilderberg, 1999).
\bibitem{Karsch} F. Karsch, Springer Lect. Notes Phys. {\bf 583}, 209 (2002).

\bibitem{Ranea} Ignacio F. Ranea-Sandoval, Sophia Han, Milva
  G. Orsaria, Gustavo G. Contrera, Fridolin Weber, Mark G. Alford,
  Phys. Rev. {\bf C 93}, 045812 (2016).

\bibitem{Spinella} W. M. Spinella, F. Weber, G. A. Contrera,
  M. G. Orsaria, EPJA {\bf 52}, 61 (2016).

\bibitem{Carvalho} S. M. de Carvalho, R. Negreiros, M. Orsaria,
  G. A. Contrera, F. Weber, W. Spinella, Phys. Rev. {\bf C 92}, 035810
  (2015).

\bibitem{Kapusta}J. I. Kapusta and C. Gale, Finite-Temperature Field
  Theory: Principles and Applications (Cambridge Monographs on
  Mathematical Physics.~ Cambridge: Cambridge University Press, 2006).
    
\bibitem{Contrera:2009hk} G. A. Contrera, D. Gomez Dumm,
  N. N. Scoccola,Phys. Rev {\bf D 81}, 054005 (2010).

\bibitem{BML} T. J. Burrows, A. Mazurek and
  J. M. Lattimer. Astrophys. J., 251:325 (1981); A. Burrows and
  J. M. Lattimer. Astrophys. J. {\bf 307}, 178 (1986).

\bibitem{Lugones:2016} G. Lugones, Eur. Phys. J.  {\bf A 52}, 53 (2016).

\end{thebibliography}
\end{document}